\documentclass[aps,pra,twocolumn]{revtex4-1}
\usepackage{amsmath,amsfonts,amssymb}
\usepackage{graphicx,float,calc}
\usepackage{color,bm}
\usepackage{ulem}
\usepackage[breaklinks,colorlinks,urlcolor=blue,citecolor=blue,linkcolor=blue]{hyperref}

\begin{document}

\title{Chiral magnetic effect in three-dimensional optical lattices}
\author{Zhen Zheng$^1$}\thanks{These authors contributed equally to this work.}
\author{Zhi Lin$^{2}$}\thanks{These authors contributed equally to this work.}
\author{Dan-Wei Zhang$^3$}\thanks{danweizhang@m.scnu.edu.cn}
\author{Shi-Liang Zhu$^{4,3}$}
\author{Z. D. Wang$^1$}\thanks{zwang@hku.hk}
\affiliation{$^1$Department of Physics and Center of Theoretical and Computational Physics, The University of Hong Kong, Pokfulam Road, Hong Kong, China}
\affiliation{$^2$School of Physics and Materials Science, Anhui University, Hefei 230601, China}
\affiliation{$^3$Guangdong Provincial Key Laboratory of Quantum Engineering and Quantum Materials, GPETR Center for Quantum Precision Measurement and SPTE, South China Normal University, Guangzhou 510006, China}
\affiliation{$^4$National Laboratory of Solid
	State Microstructures and School of Physics, Nanjing University,
	Nanjing 210093, China}
\begin{abstract}

Although Weyl semimetals have been extensively studied for exploring rich topological physics,  the direct observation of the celebrated chiral magnetic effect (CME) associated with the so-called dipolar chiral anomaly has long intrigued and challenged physicists, still remaining elusive in nature. Here we propose a feasible scheme for experimental implementation of ultracold atoms that may enable us to probe the CME with a pure topological current in an artificial Weyl semimetal.
The paired Weyl points with the dipolar chiral anomaly emerge in the presence of the well-designed spin-orbital coupling and laser-assisted tunneling. Both of the two artificial fields are readily realizable and highly tunable via current optical techniques using ultracold atoms trapped in three-dimensional optical lattices, providing a reliable way for manipulating Weyl points in the momentum-energy space. By applying a weak artificial magnetic field, the system processes an auxiliary current originated from the topology of a paired Weyl points, namely, the pure CME current.
This topological current can be extracted from measuring the center-of-mass motion of ultracold atoms,
which may pave the way to directly and unambiguously observe the CME in experiments.

\end{abstract}

\maketitle

\section{Introduction}

Quantum simulation is one of the major topics in ultracold-atom research.
Compared to condensed matter, ultracold-atomic gases trapped in optical lattices (OLs) host advantages such as precise control of the system parameters and the disorder \cite{quantum-simulation-rev-nphys,quantum-simulation-rev-sci,lewenstein2007ultracold,DWZhang2018,goldman2016topological,Cooper2019}.
Furthermore, effective gauge fields, for instance the spin-orbital coupling (SOC) \cite{lin2011spin,fermi-soc-1,fermi-soc-2,fermi-soc-3,fermi-soc-rashba,Wu83,SLZhu2006,Ruseckas2005,Osterloh2005} and the magnetic field \cite{jaksch-njp,lin2009synthetic,cold-atom-mag,laser-hop-1,laser-hop-2,celi2014synthetic,Mancini1510},
can be synthesized by means of optical techniques \cite{raman-rev-2011,raman-rev-2014}.
Thus, ultracold atoms offer a versatile platform for quantum simulation and exploration of diverse condensed-matter phenomena, such as topological quantum states of matter \cite{DWZhang2018,goldman2016topological,Cooper2019}, which have emerged as a popular topic in recent years.

Among topological materials, Weyl semimetals (WSMs) associated with Weyl points (WPs) \cite{weyl-Burkov,weyl-Kim,Xu613,huang2015weyl,Lu622,weyl-Zhao,weyl-Lv}
provide a promising avenue for investigating and understanding massless chiral fermions in the relativistic quantum field theory \cite{volovik2003universe},
and thereby attract tremendous research interest.
In WSMs, WPs stem from the fact that the conduction and valence bands contact only at discrete points in the three-dimensional (3D) Brillouin zone (BZ) \cite{graphene-rmp}.
By virtue of the broken time reversal or (and) inversion symmetry, WPs in such an electronic structure emerge in pairs \cite{dirac-sm}. The paired WPs have wide applications for topological states,
for instance the simulation of the long-sought magnetic monopoles in momentum space and the associated Fermi arc modes, whose spatial distributions are localized on surfaces of materials \cite{hasan2017discovery}.

As a result of recent progress in the investigation of WSMs, it has been revealed that there is an exotic kind of anomalous topological current in the electromagnetic response theory of WPs. This phenomenon is known as the chiral magnetic effect (CME), and arises from the topology of paired WPs \cite{cme-prd,cme-prb,cme-prl-2012,cme-prl-2013,kharzeev2014chiral}.
The emergent topological current is proportional to not only the external magnetic field, but also the energy shift between the paired WPs even in the absence of averaged electric fields in real space. Although some possible indications were reported for the CME current with other mixed currents in condensed-matter systems (e.g., usual chiral anomaly currents associated with the nonzero parallel component of electric and magnetic fields) \cite{QLi2016,Shin2017}, a direct and smoking-gun probe of pure CME currents remains elusive owing to the lack of flexible techniques of engineering and manipulating WPs in real materials. A recent experiment shows that
superconducting quantum circuits provide a possibility for mimicking CME currents in a virtual sense \cite{yuyang-weyl};
no real particle currents were detected there.
On the other hand, WPs can be engineered via the laser-assisted tunneling or synthetic SOCs for ultracold atoms in OLs \cite{JHJiang2012,cold-atom-weyl-raman,cold-atom-weyl-Zhang,song2019observation,WYHe2016,ZLi2016,BZWang2018,LJLang2017,YXu2016b,XKong2017,Shastri2017}.
By deliberately designing the laser's configurations, the band structure possesses WP pairs in the BZ.
Since ultracold atoms provide great controllability for studying topological matter \cite{DWZhang2018,goldman2016topological,Cooper2019},
it inspires us to search for a promising experimental scheme to manipulate WPs to probe the exotic CME unambiguously.

In this paper, we present a feasible proposal for simulating the CME with ultracold atoms in a 3D OL.
Our main results are as follows: (i) The paired WPs are engineered in the presence of a Rashba-type-like SOC and the band inversion with respect to spins. Here the SOC has been realized in experiments using ultracold atoms \cite{fermi-soc-rashba,Wu83}, which is our starting point, while the band inversion can be naturally introduced in the atomic operator representation.
It results in separated WPs with opposite chirality along the $z$ axis of the BZ.
The distance between the paired WPs can be tuned by the spin imbalance or additional Zeeman field.
(ii) The laser-assisted-tunneling technique \cite{cold-atom-mag} serves as a perfect tool for engineering
the energy shift between the paired WPs and the effective magnetic field.
It paves the way for engineering WPs associated with the CME.
(iii) In the generation of the energy shift and magnetic field,
their magnitudes and directions are all controllable by the laser fields.
It facilitates the observation of the topological current [cf. Eq. (\ref{eq-current})], providing direct evidence of the CME.

\section{Weyl Hamiltonian}

The system of our interest is governed by the following WSM Hamiltonian in a tight-binding model,
\begin{align}
H(\bm{k}) &= \sum_{\bm{k}} [m_z - 2t\sum_{l=x,y,z}\cos(k_l d)]\sigma_z +2\lambda \sin(k_zd)\sigma_0\notag\\
& +2\alpha \sin(k_xd)\sigma_x + 2\alpha \sin(k_yd) \sigma_y \,.
\label{eq-h-k}
\end{align}
Here we have chosen the spinor base $a_{\bm{k}}=(a_{\bm{k},\uparrow},a_{\bm{k},\downarrow})^T$,
with $a_{\bm{k},\sigma}$ denoting the annihilation operator of spin-$\sigma=\uparrow,\downarrow$ atoms.
$m_z$ is the energy shift with respect to spins. $t$ is the nearest-neighbor tunneling magnitude. $d$ is the lattice constant.
$\lambda$ characterizes a spin-independent tunneling along the $z$ direction, which will play a key role for the energy shift between the paired WPs. $\alpha$ is the SOC strength. $\sigma_{0,x,y,z}$ are Pauli matrices.

We first investigate the simple case with $\lambda=0$.
The $m_z$ term breaks the time-reversal symmetry,
while the SOC destroys the space-inversion symmetry.
However, the Hamiltonian (\ref{eq-h-k}) with $\lambda=0$ is still inversion invariant along the $z$ direction.
Hence in the BZ, the paired WPs with opposite chirality reside at $\bm{k}_W=(0,0,\pm k_W)$ on the $k_z$ axis with $k_W\equiv \arccos[(m_z-4t)/2t]$.
The chirality of each WP is also equivalent to the Chern number of a closed surface that encloses a WP in momentum space \cite{cold-atom-weyl-Zhang},
and thereby serves as a topological invariant.
Here for simplicity but without loss of generality, we have assumed that $2t<m_z<6t$.
In the vicinity of the WPs, Hamiltonian (\ref{eq-h-k}) is approximately linear with respect to $\bm{k}$,
\begin{equation}
H_\xi = v_{W\parallel} k_x\sigma_x + v_{W\parallel} k_y\sigma_y + \zeta v_{W\perp} k_z\sigma_z \,.
\label{eq-weyl-h-k-linear}
\end{equation}
We can see that the WPs resemble massless Weyl fermions with chirality $\zeta=+$($-$) for the left-(right-)handed one.
Here $v_W$'s denote the velocities of the WPs.
Such a band structure gives rise to a rich topological phenomenon.
The paired WPs can simulate the magnetic monopoles in momentum space by recognizing the WP's chirality as the magnetic charge.
Due to the broken time-reversal symmetry, each WP is separated from the other one with opposite chirality in the BZ.
As a result, the emergent Fermi arc mode, which is evidenced as the topological surface state \cite{hasan2017discovery}, is in analog to the Dirac string that connects two magnetic monopoles with opposite charges.

\begin{figure}[t]
\centering
\includegraphics[width=0.49\textwidth]{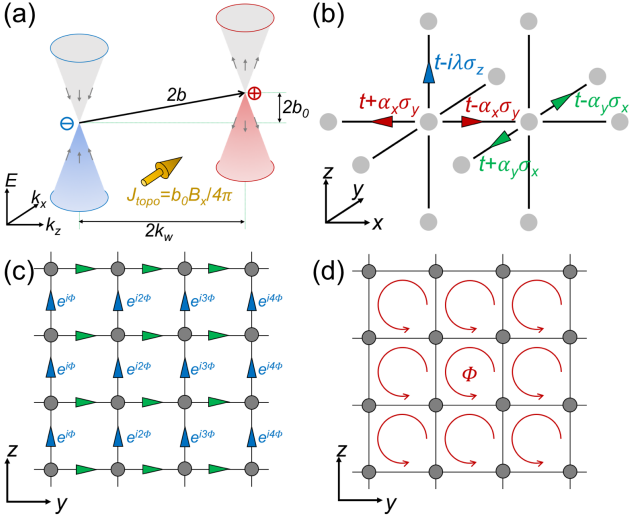}
\caption{(Color online) (a) Sketch of the simulated CME current.
(b) Sketch of the 3D lattice accompanied by SOC (red and green arrows) and spin-dependent tunneling (blue arrows).
We denote $\alpha_x=\alpha_y\equiv \alpha (-1)^{j_x+j_y+j_z}$.
(c) Raman lasers placed in the $y$ direction enable tunneling with the spatially modulated phase along the $z$ direction.
(d) Nonzero magnetic flux per square plaquette in the $y$-$z$ plane.
The engineered effective magnetic field is thereby along the $x$ direction.}
\label{fig-model}
\end{figure}

When $\lambda\neq0$, the paired WPs still reside at $\pm k_W$, while the $\lambda$ term leads to an energy shift $b_0=2\lambda\sin(k_Wd)$ between them.
In particular, the left-handed WP is separated by a four-dimensional (4D) vector $2b$ from the right-handed one in energy-momentum space, where $b= (b_0,\bm{b})$ with $\bm{b}=(0,0,k_W)$, which is shown in Fig. \ref{fig-model}(a).
In such a band structure,
if one applies an effective gauge field, an additional topological action will be introduced in terms of the Chern-Simons form: $S_\mathrm{topo}= -\frac{1}{8\pi^2}\int d^4x \, \varepsilon^{\mu\nu\lambda\sigma}b_\mu A_\nu \partial_\lambda A_\sigma$ \cite{cme-prb,yuyang-weyl,ZLin2019}.
Here $A_\nu$ denotes the 4D vector potential of the gauge field,
and $\varepsilon^{\mu\nu\lambda\sigma}$ is the Levi-Civit\`{a} symbol.
For simplicity, we assume the gauge field to be an effective magnetic field $\bm{B}$.
By varying $S_\mathrm{topo}$ with respect to $A_\nu$,
it gives rise to an intrinsic topological current (hereafter we set the charge $q=1$ for neutral atoms):
\begin{equation}
\bm{J}_\mathrm{topo}=\frac{b_0}{4\pi^2}\bm{B} \,. \label{eq-current}
\end{equation}
It implies that $\bm{J}_\mathrm{topo}$ can be regarded as a topological response to the applied magnetic field $B$ in the presence of the chirality imbalance, termed CME.
In the following, we shall focus on the engineering of a highly tunable Weyl Hamiltonian and probing a topological particle current of CME with cold atoms.

\section{Hamiltonian Engineering}

Now we present the proposal for realizing the Weyl Hamiltonian in ultracold atomic gases.
We consider a 3D OL system, in which the atomic internal states are chosen as the pseudo-spins $\uparrow\downarrow$.
In accordance with Eq. (\ref{eq-h-k}),
we start with the following Hamiltonian composed of three terms:
\begin{equation}
H = H_0 + H_z + H_\mathrm{soc} \,, \label{eq-h-start}
\end{equation}
which is illustrated in Fig. \ref{fig-model}(b).
The first term $H_0$ describes the free-particle Hamiltonian trapped by the 3D OL potential,
$V_L(\bm{r})=V_L\sum_{l=x,y,z}\sin^2(k_Ll)$, with the recoil momentum $k_L=\pi/d$.
We use natural units $\hbar=1$, unless stated otherwise.
The tight-binding form is given by
\begin{align}
H_0 &= \sum_{j,\sigma=\uparrow\downarrow} (-t a_{j+1,\sigma}^\dag a_{j,\sigma} +H.c.) \notag\\
&+\sum_{j}m_z (a_{j,\uparrow}^\dag a_{j,\uparrow}-a_{j,\downarrow}^\dag a_{j,\downarrow}) \,. \label{eq-h-xy}
\end{align}
Here, $a_{j,\sigma}$ denote the annihilation operator of spin-$\sigma$ atoms on the $j$th site.
$m_z=(\mu_\uparrow-\mu_\downarrow)/2$ with $\mu_\sigma$ denoting the chemical potential of spin-$\sigma$ atoms.
We have discarded the energy constant term $(\mu_\uparrow+\mu_\downarrow)/2$.
Along the $z$ direction, besides the spin-independent tunneling $t$,
we generate an additional spin-dependent one by means of the laser-assisted-tunneling technique.
This is attainable by a Raman transition between nearest-neighbor sites,
in which the energy offset of adjacent sites is provided by a titled magnetism and compensated by the Raman detuning with proper two-photon frequencies (see Appendix \ref{sec-app-hop}).
It gives the second term of Hamiltonian (\ref{eq-h-start}),
\begin{equation}
H_z = \sum_{j}i\lambda (a_{j+\bm{e}_z,\uparrow}^\dag a_{j,\uparrow} -
a_{j+\bm{e}_z,\downarrow}^\dag a_{j,\downarrow} ) +H.c. \label{eq-h-z}
\end{equation}
Here $\lambda$ denotes the spin-dependent tunneling magnitude.

The last term of Hamiltonian (\ref{eq-h-start}) describes the SOC:
$H_\mathrm{soc}=\int d\bm{r}d\bm{r}'\, [M(\bm{r}) \psi_\uparrow^\dag(\bm{r}) \psi_\downarrow(\bm{r}') +H.c.]$.
Here $\psi_\sigma$ ($\psi_\sigma^\dag$) is the annihilation (creation) operator of spin-$\sigma$ atoms, and
$M(\bm{r})$ denotes the coupling mode associated with a spatially modulated magnitude.
In realistic experiments of ultracold atoms,
this term can be realized via the optical dressing \cite{fermi-soc-rashba}.
In order to generate a Rashba-type-like SOC,
we can design $M(\bm{r})=i M_x(\bm{r})+M_y(\bm{r})$
with $M_x(\bm{r})=M_0\sin(k_Lx)\cos(k_Ly)\cos(k_Lz)$
and $M_y(\bm{r})=M_0\sin(k_Ly)\cos(k_Lx)\cos(k_Lz)$.
Since the Wannier wave function of atoms in lattices denoted by $W(\bm{r})$ is an even function in real space,
and $\sin(k_Ll)$ is antisymmetric with respect to the $l=x,y,z$ axis,
the on-site terms in $H_\mathrm{soc}$ are eliminated.
Thus, $H_\mathrm{soc}$ can be expressed as
\begin{align}
H_\mathrm{soc}&=\sum_j e^{i(j_x+j_y+j_z)\pi} [i\alpha ( a_{j+\bm{e}_x,\uparrow}^\dag a_{j,\downarrow} - a_{j-\bm{e}_x,\uparrow}^\dag a_{j,\downarrow} ) \notag\\
&+\alpha ( a_{j+\bm{e}_y,\uparrow}^\dag a_{j,\downarrow} - a_{j-\bm{e}_y,\uparrow}^\dag a_{j,\downarrow} ) ]
+H.c. \label{eq-h-soc}
\end{align}
Here $\alpha=\int d \bm{r} \, M_{x,y}(\bm{r})W_\uparrow^*(\bm{r}+d\bm{e}_{x,y})W_\downarrow(\bm{r})$ is the SOC strength,
and $j_l$ ($l=x,y,z$) denotes the $l$-direction component of the $j$th site.
The detailed information of Eqs. (\ref{eq-h-xy})--(\ref{eq-h-soc}) is given in Appendix \ref{sec-app-tmb}.

We make the operator representation for the spin-$\downarrow$ atoms,
$a_{j,\downarrow}\rightarrow a_{j,\downarrow}e^{i(j_x+j_y+j_z)\pi}$ that conserves the commutation (anti-commutation) of bosons (fermions).
In this representation, we obtain the Hamiltonian from Eqs. (\ref{eq-h-xy})-(\ref{eq-h-soc}),
\begin{align}
\tilde{H}&=\sum_{j} [( -ta_{j+1}^\dag \sigma_z a_{j} +i\lambda a_{j+\bm{e}_z}^\dag \sigma_0 a_{j}+H.c.)
+m_z a_{j}^\dag \sigma_z a_{j}] \notag\\
&+ \sum_{j,\eta=\pm} \eta\frac{\alpha}{2} ( i a_{j+\eta\bm{e}_x}^\dag \sigma_x a_{j}
+ a_{j+\eta\bm{e}_y}^\dag \sigma_y a_{j} +H.c.) \, \label{eq-h-site-final}
\end{align}
in the base $a_j\equiv(a_{j,\uparrow},a_{j,\downarrow})^T$.
One can see that the band inversion with respect to spins is naturally introduced in the operator representation.
Equation (\ref{eq-h-site-final}) in momentum space corresponds to the desired WSM Hamiltonian (\ref{eq-h-k}).

We hereby discuss the generation of the applied magnetic field $\bm{B}$.
In ultracold-atom experiments, effective magnetic fields acting on neutral atoms can be synthesized via the laser-assisted tunneling  \cite{laser-hop-1,laser-hop-2}.
In our proposal, this technique has been used for engineering $H_z$.
Therefore, $\bm{B}$ can be simultaneously introduced if the counter-propagating lasers are placed in the $x$-$y$ plane.
For simplicity, we assume they are placed along the $y$ direction.
A spatially modulated phase of nearest-neighbor tunneling, along the $z$ direction, can be inherited from the two-photon Raman process, as shown in Fig. \ref{fig-model}(c). In particular, the spin-dependent tunneling $t\mp i\lambda$ in $H_z$ is replaced by $(t\mp i\lambda)e^{i j_y \Phi}$, where the magnetic flux $\Phi=\delta k\cdot d$ and $\delta k$ denotes the momentum transfer in the Raman process (see Fig. \ref{fig-model}(d) and Appendix \ref{sec-app-mag}). It gives rise to an effective magnetic field $\bm{B}=B_x\approx \Phi/d^2$.

\section{Observation of CME}

The simulation of the CME is readily proposed by current techniques of ultracold bosonic or fermionic atoms.
According to Eq. (\ref{eq-current}), the topological current $\bm{J}_\mathrm{topo}$ is proportional to $\bm{B}$ and $b_0=2\lambda\sin(k_Wd)$, which are both tunable by laser parameters.
It indicates that $\bm{J}_\mathrm{topo}$ changes its direction if we change the sign of $\bm{B}$ or $\lambda$,
leading to an opposite center-of-mass (COM) motion of atoms. Thus, the topological current can be directly probed by measuring the COM density current \cite{aidelsburger2015measuring},
$\bm{J}_\mathrm{topo}=[\bm{J}(b_0)-\bm{J}(-b_0)]/2$
or $\bm{J}_\mathrm{topo}=[\bm{J}(\bm{B})-\bm{J}(-\bm{B})]/2$.
This measurement has the advantage that it cancels other effects contributed to the atomic currents and can extract the pure topological current.

\begin{figure}[t]
\centering
\includegraphics[width=0.49\textwidth]{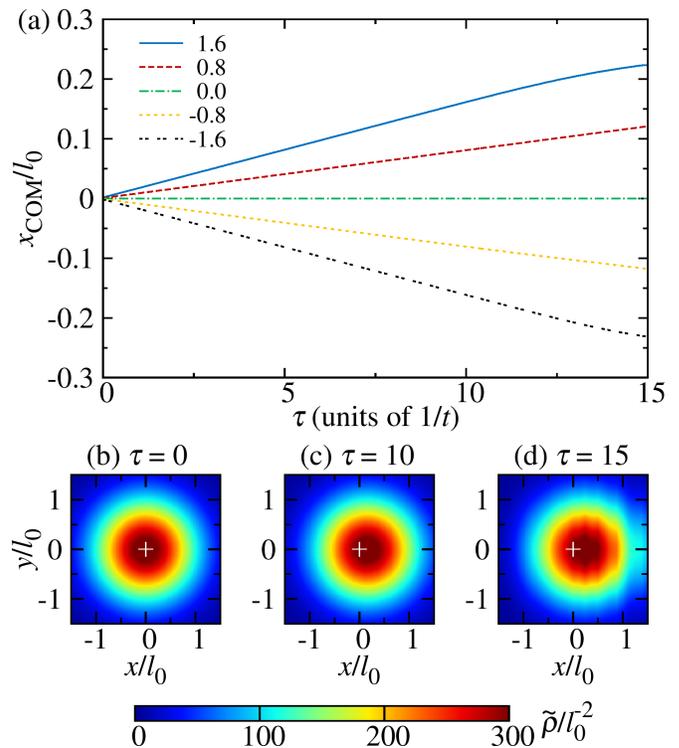}
\caption{(Color online)
(a) Time evolution of the COM position $x_{\rm COM}(\tau)$ of the atomic cloud with various velocity $v_c$.
We set $v_c$ in unit of $10^{-2}tl_0$.
(b)-(d) The density profile of the atomic cloud in the $x$-$y$ plane.
The color describes the amplitude of $\tilde{\rho}(x,y,\tau)\equiv\int \rho(\bm{r},\tau)dz$.
We set $v_c=1.6\times10^{-2}tl_0$ (blue solid line in (a)) and the total atom number $N=2000$ (yielding $l_0\sim 10^{2}d$).
The white symbol ``+" marks the trapping potential center $\bm{r}=0$.}
\label{fig-CME}
\end{figure}

We assume the atoms are prepared in an overall trapping potential, $V_{\rm trap} = \frac{1}{2}m\sum_{l=x,y,z} \omega_{\rm trap}^2 l^2$.
When the lattice constant $d\ll l_0$ ($l_0=\sqrt{\hbar/(m\omega_{\rm trap})}$),
the atomic cloud can be semiclassically recognized as a continuum system.
The initial wave packet of the atomic cloud within the trap $V_{\rm trap}$
can be given by $\psi(\bm{r},\tau=0)=\frac{1}{\mathcal{N}}\exp[-\sum_l l^2/(2 l_0^2)]$ \cite{Pethick2008BEC},
whose COM position is centered at $\bm{r}=0$.
Here, $\mathcal{N}=N^{1/2}/(\pi^{3/4}l_0^{3/2})$ is the renormalization factor ($N$ is the total atom number).
When the effective magnetic field $B_x$ turns on,
it introduces a COM velocity $v_c$, triggered by the topological current $\bm{J}_{\rm topo}$, to the wave packet.
The velocity can be approximately evaluated as $v_c\approx \bm{J}_{\rm topo}/\bar{\rho}=b_0B_x/4\pi\bar{\rho}$,
with $\bar{\rho}\approx 1/d^3$, if the lattice system is at half filling.
The hydrodynamics of the atomic density, $\rho(\bm{r},\tau)\equiv |\psi(\bm{r},\tau)|^2$,
is governed by $\partial_\tau \rho + \nabla \cdot (\rho\bm{v}_c)=0$.
In Fig. \ref{fig-CME}, we give the results of numeric simulation.
The COM position of the wave packet  $\bm{r}_{\rm COM}(\tau)\equiv\int \bm{r}\rho(\bm{r},\tau)d\bm{r}$
drifts along the $x$ direction as long as $v_c$ is nonzero.
When $v_c$ changes its sign, the wave packet evolves to a opposite direction correspondingly.
By contrast in the $y$ or $z$ direction, we find the COM of the wave packet is not shifted and statically centered at $y_{\rm COM}(\tau)=z_{\rm COM}(\tau)=0$.

\section{Experimental feasibility}

The proposed scheme can be realized by using alkaline atoms such as $^{87}$Rb \cite{Wu83} and $^{40}$K \cite{huang2016experimental} or alkaline-earth-like atoms such as $^{173}$Yb \cite{song2019observation}, in which the 2D SOC has been successfully engineered.
We briefly discuss the experimental setup with $^{40}$K as an example.
In this ultracold-atomic gas, we choose the hyperfine state $|F,m_F\rangle=|9/2,-9/2\rangle$ as pseudo-spin-$\uparrow$
and $|9/2,-7/2\rangle$ as spin-$\downarrow$.
The OL can be created by counter propagating $\lambda_L=1064$ nm lasers,
with the lattice constant $d=\lambda_L/2$ and the recoil momentum $k_L=2\pi/\lambda_L$.
The lattice recoil energy $E_R=\hbar^2k_L^2/2m\approx2\pi\hbar\times4.4$ kHz, which is chosen as the energy unit hereafter.
For specificity, we set the trap depth of the OL as $V_L=8.0E_R$.
The corresponding nearest-neighbor tunneling magnitude is $t\approx 0.091E_R$ \cite{Wannier_code}.
If we tune the laser strength $M_0=1.0\sim2.5E_R$,
the resulting SOC strength can have a range of $\alpha\approx 0.055\sim0.136E_R$. As the magnetic field is generated by the laser-assisted tunneling, the resulting magnetic flux per plaquette is obtained as $\Phi=\delta k\cdot d=\pi \lambda_L/\lambda_K$ ($\lambda_K = 2\pi/\delta k$) \cite{laser-hop-1}.
In practice, we can tune $\lambda_K$ of the order of $\lambda_K\sim10^{-2}\lambda_L$ to obtain $\Phi=B_xd^2\sim10^{-1}\pi$, which can be regarded as a perturbative external field.

In order to observe CME with the controllable COM velocity,
we provide the following two schemes:
(i) The magnetic flux $\Phi$ is determined by the momentum transfer $\delta k$ in the Raman process.
We can exchange the relative positions of the two counter propagating lasers that generate $\Phi$
or change the direction of the magnetic field gradient that is used to generate a linear potential between nearest-neighbor sites \cite{laser-hop-1}. Thus the magnetic flux will flow to an opposite direction, i.e. $\bm{B}$ changes its sign.
(ii) We impose a global $\pi$ phase to $\lambda$ in the laser-assisted tunneling.
This can be achieved by adding an additional $\pi$-phase shift to one of the counter propagating lasers \cite{laser-hop-1},
which changes the sign of $\lambda$.

\section{Conclusion}

In summary, we have proposed to simulate the Weyl Hamiltonian associated with the CME in a 3D OL.
The paired WPs are engineered by introducing SOC and laser-assisted tunneling,
both of which are realizable in current experiments using ultracold atoms.
With the tunable energy shift of WPs and artificial magnetic field, the CME current can be extracted from the COM motion of ultracold atoms. The chiral response as a manifestation of the CME can be directly observed using an atomic wave packet. The realization of our scheme allows further exploration of the topological physics of Weyl fermions.

\begin{acknowledgments}
This work was supported by the GRF (No. HKU173057/17P) and CRF (No. C6005-17G) of Hong Kong, the NKRDP of China (Grant No. 2016YFA0301800), the NSFC (Grants No. 11704367 and No. 11604103), the NSAF (Grant No. U1830111), the Key R\&D Program of Guangdong Province (Grant No. 2019B030330001), the KPST of Guangzhou (Grant No. 201804020055), and the Startup Foundation of Anhui University (Grant No. J01003310).
\end{acknowledgments}

\appendix

\section{Initial Model Hamiltonian} \label{sec-app-model}

We start with the following Hamiltonian composed of three terms:
\begin{equation}
H = H_0 + H_\mathrm{soc} + H_z \,.
\label{eq-h-scheme-real}
\end{equation}
The first term is the single-particle Hamiltonian of atoms trapped in the optical lattice,
\begin{equation}
H_0 = \int d \bm{r} \sum_{\sigma=\uparrow\downarrow} \psi_\sigma^\dag(\bm{r}) [ -\nabla^2/2m +V_L(\bm{r}) -\mu_\sigma] \psi_\sigma(\bm{r}) \,. \label{eq-app-hop-0}
\end{equation}
Here $\psi_\sigma$ denotes the annihilation operator of spin-$\sigma$ atoms.
$V_L(\bm{r})=\sum_{l=x,y,z}V_L\sin^2(k_Ll)$ is the optical lattice potential.
$k_L=\pi/d$ is the recoil momentum with $d$ as the lattice constant.
$\mu_\sigma$ is the chemical potential.
The second term describes the SOC,
\begin{equation}
H_\mathrm{soc}=\int d \bm{r} d \bm{r}' \,
M(\bm{r}) \psi_\uparrow^\dag(\bm{r}) \psi_\downarrow (\bm{r}') +H.c. \label{eq-app-soc-0}
\end{equation}
Here $M(\bm{r})$ is the coupling mode whose magnitude is spatially modulated.
The last term describes the spin-dependent tunneling along the $z$ direction, which can be generated by laser fields,
\begin{align}
H_z &= \int d \bm{r} \, [
\Omega e^{ i \phi_1}\psi_\uparrow^\dag(\bm{r}+d\cdot \bm{e}_z)\psi_\uparrow(\bm{r}) \notag\\
&+\Omega e^{ i \phi_2}\psi_\downarrow^\dag(\bm{r}+d\cdot \bm{e}_z)\psi_\downarrow(\bm{r}) ]  +H.c. \label{eq-app-hz-0}
\end{align}
Here, $\Omega$ denotes the the laser-assisted-tunneling magnitude.
$\phi_{1,2}$ denote the spin-dependent phases.
$\bm{e}_{x,y,z}$ denote the unit vectors.

\section{Tight-Binding Model} \label{sec-app-tmb}

We use the tight-binding model (TBM) and expand $\psi_\sigma$ in terms of Wannier wave function $W(\bm{r})$,
\begin{equation}
\psi_\sigma(\bm{r}) = \sum_{j,\sigma} W(\bm{r}-\bm{r}_j)a_{j,\sigma} \,.
\end{equation}
Here $a_{j,\sigma}$ denotes the annihilation operator of spin-$\sigma$ atoms on the $j$th site.
$\bm{r}_j$ is the central position of the $j$-th site.
The Hamiltonian (\ref{eq-app-hop-0}) is rewritten as
\begin{equation}
H_0= -t \sum_{j,\sigma} ( a_{j+1,\sigma}^\dag a_{j,\sigma} +H.c.) +
\sum_{j}m_z (a_{j,\uparrow}^\dag a_{j,\uparrow}-a_{j,\downarrow}^\dag a_{j,\downarrow}) \,.\label{eq-app-hop-1}
\end{equation}
Here $t$ is the nearest-neighbor tunneling magnitude, and $m_z=(\mu_\downarrow-\mu_\uparrow)/2$.
We have discarded the constant term $-(\mu_\uparrow+\mu_\downarrow)/2$.

In order to engineer a Rashba-type-like SOC, we set the coupling mode $M(\bm{r})=iM_x(\bm{r})+M_y(\bm{r})$ \cite{fermi-soc-rashba},
in which
\begin{equation}
\left\{\begin{split}
M_x(\bm{r}) = M_0\sin(k_Lx)\cos(k_Ly)\cos(k_Lz) \\
M_y(\bm{r}) = M_0\sin(k_Ly)\cos(k_Lx)\cos(k_Lz) \\
\end{split}\right. \,.
\end{equation}
Since $W(\bm{r})$ is an even function in real space, i.e., $W(\bm{r})=W(-\bm{r})$,
and $\sin(k_Ll)$ is antisymmetric with respect to the $l=x,y,z$ axis,
the on-site terms of $H_\mathrm{soc}$ will vanish.
Then Hamiltonian (\ref{eq-app-soc-0}) in the TBM is given by
\begin{align}
H_\mathrm{soc} &= \sum_j \alpha (-1)^{j_x+j_y+j_z} [ i( a_{j+ \bm{e}_x,\uparrow}^\dag a_{j,\downarrow} - a_{j- \bm{e}_x,\uparrow}^\dag a_{j,\downarrow}) \notag\\
&+ (a_{j+ \bm{e}_y,\uparrow}^\dag a_{j,\downarrow} - a_{j- \bm{e}_y,\uparrow}^\dag a_{j,\downarrow} ) +H.c.]\,, \label{eq-app-soc-1}
\end{align}
where we denote the SOC strength as $\int d \bm{r} \, M_x(\bm{r}) W^*(\bm{r}-d\cdot \bm{e}_x)W(\bm{r})=\int d \bm{r} \, M_y(\bm{r}) W^*(\bm{r}-d\cdot \bm{e}_y)W(\bm{r})\equiv\alpha$.

Hamiltonian (\ref{eq-app-hz-0}) in TBM is written as
\begin{equation}
H_z= \sum_j ( \Omega  e^{ i \phi_1} a_{j+ \bm{e}_z,\uparrow}^\dag a_{j,\uparrow} +
\Omega  e^{ i \phi_2} a_{j+ \bm{e}_z,\downarrow}^\dag a_{j,\downarrow} ) +H.c.
\end{equation}
The phases are tuned as
\begin{equation}
\phi_1 = -\phi_2 = \phi \,,
\end{equation}
and we denote
\begin{equation}
\Omega \cos(\phi) = -t \,,\qquad
\Omega \sin(\phi) = \lambda \,. \label{eq-app-denote-hz}
\end{equation}
Then $H_z$ is rewritten as
\begin{align}
H_z &= \sum_j  [ i\lambda( a_{j+ \bm{e}_z,\uparrow}^\dag a_{j,\uparrow}
- a_{j+ \bm{e}_z,\downarrow}^\dag a_{j,\downarrow}) +H.c.] \notag\\
&-t \sum_{j,\sigma} (a_{j+ \bm{e}_z,\sigma}^\dag a_{j,\sigma} + H.c.)  \,. \label{eq-app-hz-1}
\end{align}
The detailed derivations of $H_z$ are given later.
We remark that, since $H_z$ is engineered by laser fields,
the natural $z$-direction hopping $t$ in Hamiltonian (\ref{eq-app-hop-1}) is suppressed in the presence of the titled magnetism \cite{laser-hop-1}, but is restored in Eq.(\ref{eq-app-hz-1}).

We make the following operator transformation that conserves the anticommutation of fermions:
\begin{equation}
a_{j,\uparrow} \rightarrow a_{j,\uparrow} \,,\qquad
a_{j,\downarrow} \rightarrow (-1)^{j_x+j_y+j_z}a_{j,\downarrow} \,.
\end{equation}
In this representation, we obtain the effective Hamiltonian as the form of Eq.(\ref{eq-h-site-final}).

\section{Laser-Assisted Tunneling} \label{sec-app-hop}

We hereby give the detailed derivations of $H_z$.
For our proposal, we add a titled magnetic field along the $z$ direction whose magnitude is $\Delta$.
Atoms on the $j$-th and $(j+1)$-th sites are coupled via a two-photon Raman process using counter-propagating lasers,
in which the internal states are denoted as $e_{1,2}$.
In TBM, the laser-assisted tunneling Hamiltonian is given by \cite{jaksch-njp}
\begin{equation}
\mathcal{H}_z = \mathcal{H}_z^{(1)} + \mathcal{H}_z^{(2)}  \label{eq-app-raman-start}
\end{equation}
with
\begin{align}
&\mathcal{H}_z^{(1)} = \sum_{j} [ j_z\Delta a_{j,\uparrow}^\dag a_{j,\uparrow}
+j_z\Delta a_{j,\downarrow}^\dag a_{j,\downarrow} \notag\\
&+ (\Gamma_1+ j_z\Delta) e_{j,1}^\dag e_{j,1}
+ (\Gamma_2+ j_z\Delta) e_{j,2}^\dag e_{j,2}]
\end{align}
and
\begin{align}
\mathcal{H}_z^{(2)} &= \sum_j [A e^{ i(\bm{k}_1\cdot\bm{r}-\omega_1 t + \varphi_1)} e_{j,1}^\dag a_{j,\uparrow} \notag\\
&+A e^{ i(\bm{k}_1'\cdot\bm{r}-\omega_1' t + \varphi_1')} e_{j,1}^\dag a_{j+ \bm{e}_z,\uparrow} \notag\\
&+A e^{ i(\bm{k}_2\cdot\bm{r}-\omega_2 t + \varphi_2)} e_{j,2}^\dag a_{j,\downarrow} \notag\\
&+A e^{ i(\bm{k}_2'\cdot\bm{r}-\omega_2' + \varphi_2')} e_{j,2}^\dag a_{j+ \bm{e}_z,\downarrow} +H.c. ] \,.
\end{align}
Here we denote the energy levels $e_{1,2}$ as $\Gamma_1=E_{e_1}-E_{a_\uparrow}$ and $\Gamma_2=E_{e_2}-E_{a_\downarrow}$.
$A$ is the laser field strength.
In real experiments, we can tune $\omega_{1,2}=\omega_{1,2}'+\Delta$.
In order to obtain a time-independent Hamiltonian, we make the following rotation,
\begin{align}
U &= \exp\{ i\sum_j [j_z\Delta (a_{j,\uparrow}^\dag a_{j,\uparrow}+a_{j,\downarrow}^\dag a_{j,\downarrow}
+e_{j,1}^\dag e_{j,1} + e_{j,2}^\dag e_{j,2} ) \notag\\
&+(\omega_1 e_{j,1}^\dag e_{j,1} + \omega_2 e_{j,2}^\dag e_{j,2} ) ]t\}
\end{align}
We remark that, since $U$ depends solely on the index $j_z$ other than $j$,
Hamiltonians (\ref{eq-app-hop-1}) and (\ref{eq-app-soc-1}) remain unchanged under the $U$ rotation.
$\mathcal{H}_z$ is rewritten as $\mathcal{H}_z'=U \mathcal{H}_z U^\dag -  i U \partial_t U^\dag$,
whose detailed formula is as follows:
\begin{align}
\mathcal{H}_z' &= \sum_{j} \Big[ (\Gamma_1-\omega_1) e_{j,1}^\dag e_{j,1}
+ (\Gamma_2-\omega_2) e_{j,2}^\dag e_{j,2} \notag\\
&+A e^{ i (\bm{k}_1\cdot\bm{r}+ \varphi_1)} e_{j,1}^\dag a_{j,\uparrow} \notag\\
&+A e^{ i (\bm{k}_1'\cdot\bm{r}+ \varphi_1')} e_{j,1}^\dag a_{j+ \bm{e}_z,\uparrow} \notag\\
&+A e^{ i (\bm{k}_2\cdot\bm{r}+ \varphi_2)} e_{j,2}^\dag a_{j,\downarrow} \notag\\
&+A e^{ i (\bm{k}_2'\cdot\bm{r}+ \varphi_2')} e_{j,2}^\dag a_{j+ \bm{e}_z,\downarrow} +H.c. \Big] \,.
\end{align}
Adiabatically eliminating $e_{1,2}$, we obtain
\begin{align}
&\mathcal{H}_z'' = \sum_{j} \{
-\frac{A^2}{\Gamma_1-\omega_1} e^{ i [(\bm{k}_1-\bm{k}_1')\cdot\bm{r}+(\varphi_1-\varphi_1')]} a_{j+ \bm{e}_z,\uparrow}^\dag a_{j,\uparrow} \notag\\
&-\frac{A^2}{\Gamma_2-\omega_2} e^{ i [(\bm{k}_2-\bm{k}_2')\cdot\bm{r}+(\varphi_2-\varphi_2')]} a_{j+ \bm{e}_z,\downarrow}^\dag a_{j,\downarrow}
+H.c. \} \,.\label{eq-app-raman-adiabatically}
\end{align}
where we have discarded the global constant Stark shift of $a_{\sigma}$.
In real experiments, we denote $\Omega_j\equiv\Omega e^{i\cdot j \cdot \delta k\cdot d}$ with
$\bm{k}_{1,2}-\bm{k}_{1,2}' \equiv \delta\bm{k}$ and
$\frac{A^2}{\Gamma_{1,2}-\omega_{1,2}}\equiv-\Omega$.
The phases are denoted as $\varphi_{1,2}-\varphi_{1,2}'\equiv \phi_{1,2}$.
Then we obtain the spin-dependent tunneling Hamiltonian,
\begin{align}
H_z &= \sum_{j} (
\Omega_j e^{ i \phi_1} a_{j+ \bm{e}_z,\uparrow}^\dag a_{j,\uparrow} +
\Omega_j e^{ i \phi_2} a_{j+ \bm{e}_z,\downarrow}^\dag a_{j,\downarrow}
+H.c. ) \,.\label{eq-app-raman-derive-final}
\end{align}

\section{Artificial Magnetic Field} \label{sec-app-mag}

The artificial magnetic field can be simultaneously engineered by the spin-dependent tunneling given by Eq. (\ref{eq-app-raman-derive-final}).
We can place the counter-propagating lasers, for simplicity, along the $y$ direction.
Tuning $\delta k = \Phi/d$ and using Eq. (\ref{eq-app-denote-hz}), we have
\begin{align}
H_z &= \sum_{j} e^{ i j_y \Phi}\times [-t \sum_{\sigma} a_{j+ \bm{e}_z,\sigma}^\dag a_{j,\sigma} \notag\\
&+ i\lambda( a_{j+ \bm{e}_z,\uparrow}^\dag a_{j,\uparrow} -
a_{j+ \bm{e}_z,\downarrow}^\dag a_{j,\downarrow}) ] +H.c. \label{eq-app-mag}
\end{align}
Hamiltonian (\ref{eq-app-mag}) reveals that an emergent magnetic flux is around each plaquette in the $y$-$z$ plane,
which is illustrated in Fig. \ref{fig-model}(c) and \ref{fig-model}(d).
The flux for spin-$\uparrow\downarrow$ atoms is the same, i.e., $\Phi_\uparrow=\Phi_\downarrow=\Phi = \delta k\times d$,
which is termed the Peierls phase \cite{Peierls-phase}.
The magnitude of the artificial magnetic field can be calculated by $B\approx \hbar\Phi/d^2$,
if we assume the atomic ``charge" $q=1$ (though ultracold atoms are electronic neutral).

In this way, we have engineered a magnetic field along the $x$ direction.
If the counter propagating lasers are placed along the $x$ direction, we can obtain a magnetic field along the $y$ direction.
In summary, this scheme can realize magnetic fields whose direction is parallel to the $x$-$y$ plane.

\section{Weyl Hamiltonian} \label{sec-app-weyl}

In Eq. (\ref{eq-h-k}),
the Weyl points emerge at $\bm{k}_W=(0,0,\pm k_W)$ with $k_W=\arccos(m_z/2t-2)/d$.
At $k\approx k_W$, using $\bm{k}=\hat{\bm{k}}+\bm{k}_W$, we have
\begin{align}
H_{W+} &\approx 2t d \sin(k_Wd) \hat{k}_z \sigma_z
+ 2\lambda d\cos(k_Wd) \hat{k}_z\sigma_0 \notag\\
&+ 2\lambda \sin(k_Wd)\sigma_0
+2\alpha d\hat{k}_x\sigma_x + 2\alpha d\hat{k}_y\sigma_y \notag\\
&\equiv v_{W\perp} \hat{k}_z \sigma_z
+ v_z \hat{k}_z\sigma_0 + b_0\sigma_0 \notag\\
&+v_{W\parallel}\hat{k}_x\sigma_x + v_{W\parallel}\hat{k}_y\sigma_y \,,
\end{align}
where
\begin{align}
&v_{W\perp} =  2t d \sin(k_Wd) \,,\quad
v_{W\parallel} = 2\alpha d \,,\\
&v_z = 2\lambda d\cos(k_Wd) \,,\quad
b_0 = 2\lambda \sin(k_Wd) \,.
\end{align}
Likewise, at $k\approx -k_W$, using $\bm{k}=\hat{\bm{k}}-\bm{k}_W$, we have
\begin{align}
H_{W-} &\approx -v_{W\perp} \hat{k}_z \sigma_z
+ v_z \hat{k}_z\sigma_0 - b_0\sigma_0 \notag\\
&+v_{W\parallel}\hat{k}_x\sigma_x + v_{W\parallel}\hat{k}_y\sigma_y \,.
\end{align}
For simplicity, we assume $v_{W\perp}=v_{W\parallel}=v_W$.
The Weyl Hamiltonian is written as
\begin{equation}
H_W(\bm{k}) = v_W (k_x\sigma_x + k_y\sigma_y \pm k_z \sigma_z) + v_zk_z \pm b_0 \,.
\label{eq-app-weyl-h}
\end{equation}
In Hamiltonian (\ref{eq-app-weyl-h}), the term $v_zk_z$ destroys the symmetry of Weyl points around $\pm\bm{k}_W$.
When $m_z=2t$ or $4t$, we obtain $|k_Wd|=\pi/2$ and $v_z=0$.
Thus the Weyl points emerge at $k_z=\pm\pi/2d$.
At this time, we have $v_W=2td$ and $b_0=2\lambda$.

\bibliographystyle{apsrev4-1}
\bibliography{references}

\end{document}